\documentclass[11pt]{article}
\usepackage{graphicx}

\newcommand{\diag}{\mbox{diag}}

\title{Ellipsoid Method for Linear Programming made simple}

\author{Sanjeev Saxena\thanks{E-mail: ssax@iitk.ac.in}\\
Dept. of Computer Science and
Engineering,\\ Indian Institute of Technology,\\
Kanpur, INDIA-208 016}

\date{\today}

\begin{document}
\maketitle

\subsection*{\centering{Abstract}}

In this paper, ellipsoid method for linear programming is derived using
only minimal knowledge of algebra and matrices. 
Unfortunately, most authors
first describe the algorithm, then later prove its correctness, which
requires a good knowledge of linear algebra.

\section{Introduction}

Ellipsoid method was perhaps the first polynomial time method for linear
programming\cite{Khi}. However, it is hardly ever covered  in Computer Science
courses. In fact, many of the existing descriptions\cite{BT,PS,Kar}
first describe the algorithm, then later prove its correctness. Moreover,
to understand one 
require a good knowledge of linear algebra (like properties of
semi-definite matrices and Jacobean)\cite{BT,PS,Kar,FR}. In this paper,
ellipsoid method for linear programming is derived using only minimal
knowledge of algebra and matrices.

We are given a set of linear equations
$$Ax\geq B$$
and have to find a feasible point.
Ellipsoid method can check whether the system 
$Ax\geq B$ has a solution 
or not, and find a solution if one is present.

The algorithm generates a sequence of ellipsoids\cite{BT},
$E_0,E_1,{\ldots} ,$ with centres $x_0,x_1,{\ldots} $ such that the
solution space (if there is a feasible solution) is is inside each of
these ellipsoids. If the centre $x_i$ of the (current) ellipsoid is not
feasible, then some constraint (say) $a^Tx\geq b$ is violated (for some
row $a$ of $A$), i.e., $a^Tx_i<b$. As all points of solution space
satisfy the constraint $a^Tx\geq b$, we may add a new constraint $a^Tx>
a^Tx_i$, without changing the solution space.  
{\em{Observe that this ``added'' half-plane will also pass through $x_i$,
the centre of the ellipsoid}}. Thus, the solution space is contained in
half-ellipsoid (intersection of Ellipsoid $E_t$ with the half-plane). The
next ellipsoid $E_{i+1}$ will cover this half-ellipsoid and its volume
will be a fraction of the volume of ellipsoid $E_i$.

The process is repeated, until we find a centre $x_k$ for which
$Ax_k\geq B$ 
or until the volume of ellipsoid becomes so small that we can conclude
that there is no feasible solution.

If the set has a solution, then there is a number $U$
\cite{BT,Kar,MS,Sai}, such that each $x_i<U$, as a result, $\sum
x_i^2<nU^2$. If we scale each $x_i$, $x'_i=x_i/\sqrt{nU^2}$, then we know
that feasible point $x'$ will be inside unit sphere (centred at origin)
$\sum x_i^2\leq 1$. 

Thus, we can take the initial ``ellipsoid'' to be unit sphere centred at
origin.

\section{Special Case: $x_1\geq 0$}

Let us first assume that the added constraint is just 
$x_1\geq 0$. 

\begin{figure}
\includegraphics[height=3.0in]{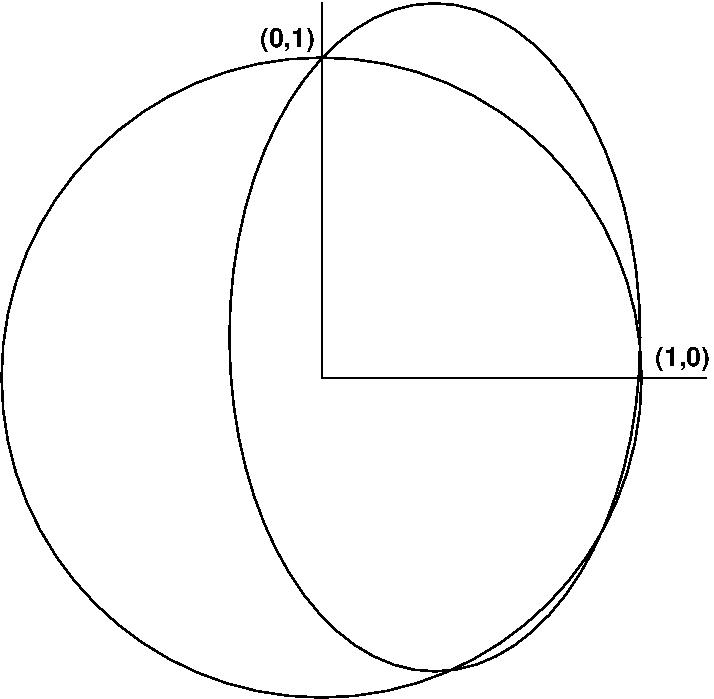}
\end{figure}

The ellipsoid (see figure), will pass through the point
$(1,0,{\ldots} ,0,0)$ and also points on (lower dimensional) sphere
$x_2^2+x_3^2+{\ldots} +x_n^2=1$ (intersection of unit sphere with
hyper-plane $x_1=0$). By symmetry, the equation of the ellipsoid should
be:

$$\alpha(x_1-c)^2+\beta(x_2^2+x_3^2+{\ldots} +x_n^2)\leq 1$$

As $(1,0,{\ldots} ,0)$ lies on the ellipsoid, $\alpha (1-c)^2=1$ or
$\alpha=\frac{1}{(1-c)^2}$. Again, the ellipsoid should also pass through
lower dimensional sphere $\sum_{i=2}^n x_i^2=1$ (with
$x_1=1$)\footnote{\sf{e.g. the point $(0,1,0,{\ldots} ,0)$ lies on
ellipsoid}}, we have $\alpha c^2+\beta=1$ or $\beta=1-\alpha
c^2=1-\frac{c^2}{(1-c)^2}=\frac{1-2c}{(1-c)^2}$. Thus, the equation
becomes
$$\frac{1}{(1-c)^2}(x_1-c)^2+\frac{1-2c}{(1-c)^2}(x_2^2+x_3^2+{\ldots}
+x_n^2)\le 1$$

or equivalently,

$$\frac{1-2c}{(1-c)^2}\sum_{i=1}^n x_i^2
   +\frac{x_1^2+c^2-x_1}{(1-c)^2}\leq 1$$

For this to be ellipsoid,  $0\leq c<\frac12$.

We want our ellipsoid to contain half-sphere. If point is inside the
half-sphere then, $\sum x'^2_i\leq 1$. Moreover, the term
$(x'^2_1-x'_1)=x'_1(x'_1-1)$ will be negative, hence, 
\begin{eqnarray*}
\frac{1-2c}{(1-c)^2}\sum_{i=1}^n x'^2_i+\frac{x'^2_1+c^2-x'_1}{(1-c)^2} &\leq&
   \frac{1-2c}{(1-c)^2}+ \frac{x'^2_1+c^2-x'_1}{(1-c)^2}\\
&\leq& \frac{1-2c}{(1-c)^2}+ \frac{c^2}{(1-c)^2}
=1
\end{eqnarray*}
Or $x'$ is also on the ellipsoid. Thus, entire half-sphere is inside our
ellipsoid.

We have certain freedom in choosing the ellipsoid (unit sphere is also an
ellipsoid which satisfies these conditions with $c=0$). We will choose
$c$ so as to reduce volume by at least a constant factor.

If we scale the coordinates as follows: 
$$x'_1=\frac{x_1}{1-c} \mbox{ and } x'_i=\frac{\sqrt{1-2c}}{1-c}x_i \mbox{ 
for }i\geq 2$$, 

Then the coordinates $x'_i$s will lie on a sphere of unit radius. Thus,
the volume of our ellipsoid will be a fraction
$\frac{1}{1-c}\left(\frac{\sqrt{1-2c}}{1-c}\right)^{n-1}$ of the volume
of a unit sphere. Or the ratio of volumes 
$$\frac{V_0}{V_1} = \frac{1}{1-c}\left(\frac{\sqrt{1-2c}}{1-c}\right)^{n-1}
 \frac{1}{1-c} \left(1-\frac{c^2}{(1-c)^2}\right)^{(n-1)/2}$$
We want $\frac{V_0}{V_1}>\alpha> 1$ (for some constant $\alpha$) or
equivalently, 
\begin{eqnarray*}
1 < \alpha &\approx& \frac{1}{1-c} \left(1-\frac{c^2}{(1-c)^2}\right)^{(n-1)/2} \mbox{
using
binomial theorem }\\
& \approx& \frac{1}{1-c}\left(1-\frac{n-1}{2}\frac{c^2}{(1-c)^2}\right) 
\approx \frac{1}{1-c}\left(1-\frac{n}{2}c^2\right) 
\approx (1+c)\left(1-\frac{nc^2}{2}\right) 
\approx 1+c-\frac{nc^2}{2} \\
\end{eqnarray*}
We should have $\frac{nc^2}{2}<c$ or $c=\Theta(1/n)$. We choose
$c=1/(n+1)$, as this maximises the ratio.\footnote{\sf{Let $f(c)=
\frac{1}{(1-c)^n}\left(\sqrt{1-2c}\right)^{n-1}=(1-c)^{-n}(1-2c)^{(n-1)/2}$\\
Then $f'(c) =(-1)(-n)(1-c)^{-(n+1)}(1-2c)^{(n-1)/2}
+(1-c)^{-n}(-2)((n-1)/2)(1-2c)^{(n-3)/2}\\
=(1-c)^{-(n+1)}(1-2c)^{(n-3)/2}\left(
n(1-2c)-(n-1)(1-c)\right)$, putting $f'(c)=0$, we get\\
$n(1-2c)=(n-1)(1-c)=n(1-c)-(1-c)$ or $nc=1-c$ or $c=1/(n+1)$}}

With this choice, 
$$\frac{V_0}{V_1} = \frac{1}{1-c} \left(1-\frac{c^2}{(1-c)^2}\right)^{(n-1)/2}
= \frac{n+1}{n}\left(1-\frac{1}{n^2}\right)^{(n-1)/2}$$
Or equivalently, using 
$1-\alpha< e^{-\alpha}$ and $1+\alpha< e^{\alpha}$
\begin{eqnarray*}
\frac{V_1}{V_0} &=& \frac{n}{n+1}\left(\frac{n^2}{n^2-1}\right)^{(n-1)/2}\\
&=& \left(1-\frac{1}{n+1}\right)\left(1+\frac{1}{n^2-1}\right)^{(n-1)/2}\\
&<& \exp\left(-\frac{1}{n+1}\right)\exp\left(\frac{1}{n^2-1}\times
\frac{n-1}{2}\right)\\
&=& \exp\left(-\frac{1}{2(n+1)}\right)\\
\end{eqnarray*}

Thus, volume decreases by a constant factor.

\section{General Case}

Let us choose our coordinate system such that the ellipsoid is aligned
with our coordinate axes. This involves rotation of axes (and possibly
translation of origin). Next, we scale the axes, such that the scaled
ellipsoid is a unit sphere (with centre as origin). Constraint $a^Tx>
a^Tx_i$ (which will be modified as axes are rotated) can again be
transformed by rotation to $x_1>0$. 

Under the new scales, ratio of the two volumes, will again be 
$\exp\left(-\frac{1}{2(n+1)}\right)$. But, as both $V_0$ and $V_1$ will
scale by the same amount, the ratio will be independent of the scale.

Thus after $k$ iterations, the ratio of final ellipsoid to initial sphere
will be at most $\exp\left(-\frac{k}{2(n+1)}\right)$. If we are to stop
as soon as volume of ellipsoid becomes less than $\epsilon$, then we want
$\exp\left(-\frac{k}{2(n+1)}\right)=\frac{\epsilon}{V_0}$, or taking
logs, $\ln \frac{V_0}{\epsilon}=\frac{k}{2(n+1)}$ or $k=O\left(n\ln
\frac{V_0}{\epsilon}\right)$.

\section{Details and Algorithm}

General axes-aligned ellipsoid with centre as $(c_1,{\ldots} ,c_n)$ is
described by

$$\frac{(x_1-c_1)^2}{a_1^2}+\frac{(x_2-c_2)^2}{a_2^2}+{\ldots}
+\frac{(x_n-c_n)^2}{a_n^2}=1$$

This can be written as $(x-c)^TD^{-2}(x-c)=1$ where $D$ is a diagonal
matrix with $d_{ii}=a_i$. If we change the origin to $c$, (using the
transformation $x'=x-c$), the equation becomes, $x'^TD^{-2}x'=1$. If we 
rotate the axes, and if $R$ is the rotation matrix $x'=Rx''$, we get
$x''^TR^TD^{-2}Rx''=1$.  This is the equation of general ellipsoid with
centre as origin. If we now wish to further rotate the axes (say to make
a particular direction as $x_1$-axis), say with rotation matrix $S$,
$x''=Sx'''$, the equation becomes $x'''^TS^TR^TD^{-2}RSx'''=1$. If we
want the centre to be $c'=(c'_1,c'_2,{\ldots} ,c'_n)$, (using the
transform $x''''-c'=x'''$), then the equation becomes (dropping all
primes) $(x-c)^T(SR)^TD^{-2}(SR)(x-c)=1$. This is the the general
equation of ellipsoid with arbitrary centre and arbitrary direction as
$x_1$-axis.

Conversely, if we are given a general ellipsoid $(x-c)^TK^{-1}(x-c)=1$
with $K^{-1}=R^TS^TD^{-2}SR$ having centre $c$, we can transform it to
unit sphere centred at origin by proceeding in reverse direction. In
other words, we first translate the coordinate system (change the origin)
using transform $x'=x-c$ to make the centre as the origin. Now the
ellipsoid is of the form $x'^T(R^TS^TD^{-2}SR)x'=1$. Next, we rotate the
coordinates twice, $x''=SRx'$, not just to make it axes-aligned, but also
to make a particular ``direction'' as $x_1$-axis. The equation becomes,
$x''^TD^{-2}x''=1$. We then scale the coordinates, $x'''=x''_1/a_i$ (or
equivalently, $x'''=D^{-1}x''$) and we are left with unit sphere
$x'''^Tx'''=1$ with centre as origin.  Thus to summarise,
$x'''=D^{-1}x''=D^{-1}SRx'=D^{-1}SR(x-c)$

{\noindent{Remark: }} Observe that as $R$ and $S$ are rotation
matrices\footnote{\sf{If $S$ is a rotation matrix, then as rotation
preserves distance,
$(x-y)(x-y)^T=((x-y)S)((x-y)S)^T=((x-y)S)(S^T(x-y)^T)=(x-y)(SS^T)(x-y)^T$,
it follows that $SS^T=I$.}} $R^T=R^{-1}$ and $S^T=S^{-1}$ $$K=(K^{-1})^{-1}
=(R^TS^TD^{-2}RS)^{-1}=R^{-1}S^{-1}D^2(S^T)^{-1}(R^T)^{-1} =
R^TS^TD^2(SR)$$

Let us now look at our constraint $a^Tx>a^T x_i$.  As $x_i=c$ the
constraint is actually, $a^T(x-c)>0$. Let $e_1=(1,0,{\ldots} ,0)^T$.
Then, in the new coordinate system, the constraint $a^T(x-c)>0$ becomes
$e^T_1x'''>0$ or $e^T_1(D^{-1}SR(x-c))=0$. Thus, we choose the direction
(or rotation matrix $S$) so that $a^T=\alpha e^T_1(D^{-1}SR)$ for some
constant $\alpha$. Or equivalently, $\alpha e_1^T= a^TR^TS^TD$, or
$\alpha e_1=DSRa$. To determine the constant $\alpha$, observe that
$\alpha^2=(\alpha e_1)^T(\alpha
e_1)=(a^TR^TS^TD)(DSRa)=a^T(R^TS^TD^2SR)a=a^TKa$, thus
$\alpha=\sqrt{a^TKa}$

We know that the next ellipsoid in the new coordinate system will be
(dropping all primes)\\
$\frac{1}{(1-c)^2}(x_1-c)^2+\frac{1-2c}{(1-c)^2}(x_2^2+x_3^2+{\ldots}
+x_n^2)=1$, or putting $c=1/(n+1)$, and 
simplifying,

$$\left(\frac{n+1}{n}\right)^2\left(x_1-\frac{1}{n+1}\right)^2
+\frac{n^2-1}{n^2} (x_2^2+x_3^2+{\ldots} +x_n^2)=1$$

Thus, the
$$K^{-1}_{\mbox{new}}=\diag\left(\left(\frac{n+1}{n}\right)^2,
\frac{n^2-1}{n^2}, \ldots, \frac{n^2-1}{n^2}\right)$$

And the new ellipsoid (in new coordinate system) can be written as (with
primes)
$$\left(x'''-\frac{e_1}{n+1}\right)^TK^{-1}_{\mbox{new}}
\left(x'''-\frac{e_1}{n+1}\right)\leq 1$$

Now, as $x'''=D^{-1}SR(x-c)$
\begin{eqnarray*}
x'''-\frac{e_1}{n+1} &=& D^{-1}SR(x-c)-\frac{e_1}{n+1}\\
&=& D^{-1}SR\left(x-c-R^{-1}S^{-1}D\frac{e_1}{n+1}\right)\\
\end{eqnarray*}
But, as $DSR a= \alpha e_1$, $e_1=\frac{1}{\alpha}DSR a$, and recalling
that $R$ and $S$ are rotation matrices, 
we have
\begin{eqnarray*}
x'''-\frac{e_1}{n+1} &=& D^{-1}SR\left(x-c-R^{-1}S^{-1}D\frac{e_1}{n+1}\right)\\
&=& D^{-1}SR\left(x-c-R^{-1}S^{-1}D\frac{1}{n+1}\frac{1}{\alpha}DSRa\right)\\
&=& D^{-1}SR\left(x-c-\frac{1}{\alpha(n+1)}(R^TS^TD^2SR)a\right)\\
&=& D^{-1}SR\left(x-c-\frac{1}{\alpha(n+1)}Ka\right)\\
\end{eqnarray*}
Or the ellipsoid is
\begin{eqnarray*}
1&\geq& \left(x'''-\frac{e_1}{n+1}\right)^TK^{-1}_{\mbox{new}}
\left(x'''-\frac{e_1}{n+1})\right)\\
&=& \left(D^{-1}SR\left(x-c-\frac{1}{\alpha(n+1)}Ka\right)\right)^T
K^{-1}_{\mbox{new}}\left(D^{-1}SR\left(x-c-\frac{1}{\alpha(n+1)}Ka\right)\right)\\
&=&
\left(x-c-\frac{1}{\alpha(n+1)}Ka\right)^TR^TS^TD^{-1}K^{-1}_{\mbox{new}}
D^{-1}SR\left(x-c-\frac{1}{\alpha(n+1)}Ka\right)\\
\end{eqnarray*}
Thus, in the original coordinate system, the new centre is:
$$c+\frac{1}{\alpha(n+1)}Ka=c+\frac{1}{n+1}\frac{Ka}{\sqrt{a^TKa}}$$

And the new matrix $K^{-1}$ is:
$K'^{-1}=R^TS^TD^{-1}K^{-1}_{\mbox{new}} D^{-1}SR$,
or equivalently,
$$K'=\left(K'^{-1}\right)^{-1}=R^{-1}S^{-1}DK_{\mbox{new}}D(S^T)^{-1}(R^T)^{-1}
=R^TS^TDK_{\mbox{new}}DSR$$

Now, matrix
$K_{\mbox{new}}=\diag\left(\frac{n^2}{(n+1)^2},\frac{n^2}{n^2-1},{\ldots}
, \frac{n^2}{n^2-1}\right)=\frac{n^2}{n^2-1}I+\left(\frac{n^2}{(n+1)^2}
-\frac{n^2}{n^2-1}\right)E_1$

Here $I$ is an identity matrix, and $E_1=\diag(1,0,{\ldots} ,0)$ is a
square matrix with only first entry as $1$ and rest as $0$.
Observe that 
\begin{eqnarray*}
E_1 &=& e_1e_1^T 
=\left(\frac{1}{\alpha}DSR a\right)\left(\frac{1}{\alpha}DSR
a\right)^T 
=\frac{1}{\alpha^2} DSRaa^TR^TS^TD
\end{eqnarray*}
And $\frac{n^2}{(n+1)^2}
-\frac{n^2}{n^2-1}
=-2\frac{n^2}{(n+1)(n^2-1)}$

Thus,
\begin{eqnarray*}
K'&=& R^TS^TD\left(\frac{n^2}{n^2-1}I-2\frac{n^2}{(n+1)(n^2-1)}
\frac{1}{\alpha^2} DSR aa^T R^TS^TD\right)DSR\\
&=& \frac{n^2}{(n^2-1)}R^TS^TDIDSR-
2\frac{n^2}{\alpha^2(n+1)(n^2-1)}R^TS^TD(DSRaa^TR^TS^TD)DSR\\
&=& \frac{n^2}{(n^2-1)}K
-2\frac{n^2}{\alpha^2(n+1)(n^2-1)}(R^TS^TD^2SR)(aa^T)(R^TS^TD^2SR)\\
&=&
\frac{n^2}{n^2-1}K-2\frac{n^2}{\alpha^2(n+1)(n^2-1)}K(aa^T)K\\
&=& \frac{n^2}{n^2-1}\left(K-\frac{2}{\alpha^2(n+1)}(Kaa^TK^T)\right)
\end{eqnarray*}	  

Or,
$$K'=\frac{n^2}{n^2-1}\left(K-\frac{2}{\alpha^2(n+1)}(Kaa^TK^T)\right)
=\frac{n^2}{n^2-1}\left(K-\frac{2}{n+1}\frac{Kaa^TK^T}{a^TKa}\right)
$$

\section{Formal Algorithm}

Observe that computation of new centre and the new ``$K$'' matrix does
not require knowledge of $S$, $R$ or $D$ matrices. In fact even the
matrix $K^{-1}$ is not required. 

After possible scaling, we can assume that the solution, if present, is
in the unit sphere centred at origin.
Thus, we initialise $K=I$ and $c=0$ 

We repeat  following two steps,
till we either get a solution, or volume of solution space becomes
sufficiently small:
\begin{enumerate}
\item Let $ax\le b$ be the first constraint for which $ac>b$. If there is
no such constraint, then $c$ is a feasible point, and we can return.

\item Make
$$c=c+\frac{1}{n+1}\frac{Ka}{\sqrt{a^TKa}}$$
and

$$K=\frac{n^2}{n^2-1}\left(K-\frac{2}{n+1}\frac{Kaa^TK^T}{a^TKa}\right)$$
\end{enumerate}

Clearly, each iteration takes $O(n^3)$ time.
Note that we do {\em{not}} explicitly construct any ellipsoids.

\bibliography{general}

\end{document}